\documentclass[multphys,vecphys]{svmult}
\usepackage{graphicx}
\begin{document}

\title*{Radio Properties of Cavities in the ICM: Imprints of AGN Activity }
\author{L. B\^{\i}rzan\inst{1},B. R. McNamara\inst{2},C. L. Carilli\inst{3}, P. E. J. Nulsen\inst{4}\and
M. W. Wise\inst{5}}
\institute{Ohio University
\texttt{birzan@helios.phy.ohiou.edu}
\and University of Waterloo \texttt{mcnamara@sciborg.uwaterloo.ca}
\and NRAO
\and Harvard-Smithsonian Center for Astrophysics
\and University of Amsterdam}
\maketitle
\section{Introduction}
\label{sec:Intr}
The cooling time of the intracluster gas in the cores of many galaxy clusters is shorter than 1 Gyr. In the absence of heating a ``cooling flow" is expected to form, but \emph{Chandra}, ASCA and \emph{XMM-Newton} spectra do not show the expected signature of cooling  below 2 keV \cite{pete01}. At the same time, \emph{Chandra} images have shown that radio jets interact with the intracluster medium (ICM) and are energetically able to offset the radiative losses in many systems \cite{birz04, raff06}. In recent years it has also been recognized that active galactic nuclei (AGN) may prevent the formation of extremely bright galaxies \cite{bens03, bowe06}. 

Motivated by the impact that AGN have on large scale structure formation and on the ICM, we investigate the properties of AGN in cluster cores. We have constructed a radio data set at 327 MHz, 1400 MHz, 4500 MHz and 8500 MHz using the Verry Large Array (VLA). We show that 327 MHz is a better tool than higher frequencies for studding  the history of AGN activity in the cores of clusters over the past several hundred million years.
   
\section{Sample}
\label{sec:Samp}
We present a sample of 16 objects taken from the samples of \cite{birz04, raff06} of systems with well-defined surface brightness depressions associated with their radio sources. Our sample consist of 15 galaxy clusters and one elliptical galaxy, M84\index{M84}. There is a broad range in redshift from 0.0035 (M84\index{M84}) to 0.35 (RBS 797\index{RBS797}). 

\section{Data Reduction and Analysis}

\subsection{Radio Observations, Data Reduction and Radio Analysis}
In order to have data at 327 MHZ, 1.4 GHz, 4.5 GHz  and 8.5 GHz we use the VLA archive and new measurements made when the archive data was insufficient. The new observations date from December 2004 to December 2006.  Using only the flux in the lobes and by fitting the data with a power law electron injection model, we computed the break frequency ($\nu_{\rm{C}}$) above which the spectrum steepens. We computed the bolometric radio luminosity by integrating the spectrum (calibrated at 327 MHz) between 10 MHZ and 10 GHZ, using the spectral index between 327 MHz and 1400 MHz \cite{birz04}. We also computed the synchrotron ages, which depend on the break frequency and the magnetic field: $t_{\rm{syn}} \propto B^{-3/2}\nu_{\rm{C}}{-1/2}$ \cite{alex87}.

\subsection{X-ray Analysis}
The total energy required to create the X-ray cavities is estimated as the work done by the jets against the ICM plus the inernal energy \cite{birz04}:
$E_{\rm{cav}}=pV+(\gamma-1)^{-1}pV$,  where $p$ is the gas pressure, $V$ is the cavity volume and  $ \gamma,$ the ratio of the gas specific heats, is assumed to be 4/3 (relativistic gas). The cavities' sizes and positions were measured from the X-ray images and the pressures are those at the cavities' centers. The age for each cavity ($t_{\rm{cav}}$) was calculated in three different ways: the sound speed age, buoyancy age  and refill age  \cite{birz04}.  The cavity power is defined as: $P_{\rm{cav}}=E_{\rm{cav}}/t_{\rm{cav}}$.  The X-ray properties that we used in this work are from \cite{raff06}.

\section{Results and Discussion}
\subsection{Radiative Efficiencies}
An important problem is the degree of coupling between jet power (traced by the cavities) and synchrotron power (the bolometric radio luminosity). This coupling was discussed in detail in  \cite{birz04}, where we plotted jet power ($P_{\rm{cav}}$) versus synchrotron power. In that paper, we used data from the literature to calculate the bolometric radio luminosity. Figure 1 is an upgraded version of Figure 1 from \cite{birz04}. Here we analyzed all of the radio sources in a consistent way in order to calculate the bolometric radio luminosity. For consistency, we separate radio filled cavities from radio ghost cavities as we did in our previous work \cite{birz04}.  

Figure 1 shows the same trend between bolometric radio luminosity and cavity power as in \cite{birz04}. The most luminous radio objects generally have the largest cavity powers. This trend is shared again by both radio filled cavities and radio ghost cavities. However, there is a great deal of scatter in this relation, similar to our previous finding. We conclude that the scatter is intrinsic to the radio data, for reasons that include radio aging and adiabatic expansion.
 
Figure 1 differs from our previous finding only in the details, but we see again that the ratio of cavity power to radio power ranges from a few to a  
 \begin{figure}
\includegraphics[width=5cm]{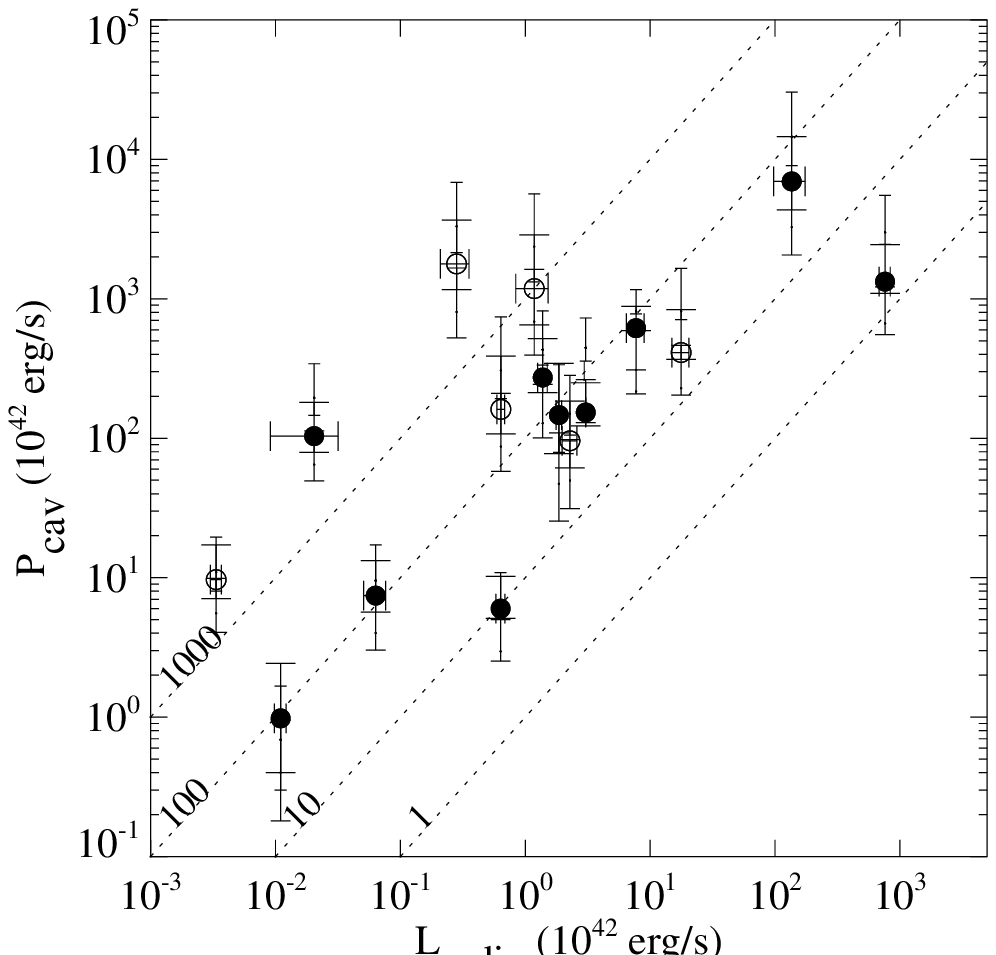}
\includegraphics[width=5cm]{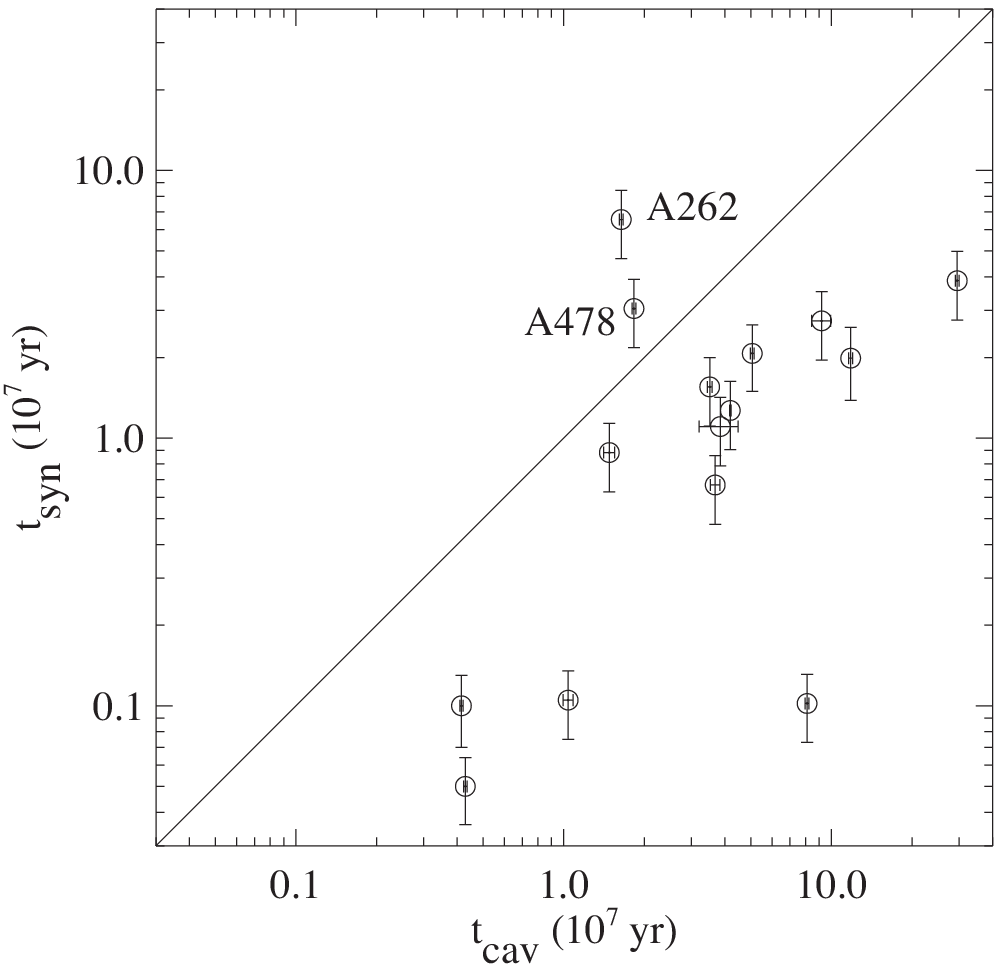}
\caption{\emph{Left}: Cavity power versus total bolometric radio power. The errors in cavity power reflect the range in cavity ages \cite{birz04}. Filled symbols denote radio-filled cavities and open symbols denote ghost cavities. The lines denotes different ratios of cavity power to bolometric radio luminosity. \emph{Right}: Synchrotron age versus buoyancy age for the entire sample.  }
\end{figure} 
few thousand. Theoretical arguments (e.g., \cite{bick97}) predict that the ratio of jet power to radio power should range between $10-100$. Most objects in our sample fall within this range, with the exception of A1835\index{A1835}, A262\index{A262}, A478\index{A478} and RBS 797\index{RBS797}, which are very inefficient radiators (radiative efficiencies of $\sim$0.001). We don't see a clear separation between the radio filled cavities and the ghost cavities, but in general the ghost cavities have a higher ratio of cavity power to radio power (i.e. they are less efficient). 

We conclude that the scatter that we see in Figure 1 is intrinsic to the radio data. It is important to note that we didn't include the contribution of shocks in the cavity power calculation for any of the objects (e.g., Hydra A\index{HydraA}  \cite{wise06} and M87\index{M87}  \cite{form06}, both presented during this conference). In cases where shocks are present, the cavity powers are lower limits to the jet powers, and as a consequence the radiative efficiencies are overestimated.

 \subsection{Age Estimates}

Using both X-ray and radio data, we can compare the synchrotron ages of the radio lobes to the buoyancy and sound speed ages of the cavities. The synchrotron age depends on the break frequency, which was derived using multifrequency radio data, and the magnetic field strength. In order to calculate the magnetic field we assumed equipartition, equal energy in protons and electrons ($k=1$), and that relativistic particles and magnetic fields occupy the same volume (filling factor $\phi=1$).  Most of the objects range between $t_{\rm{syn}}=t_{\rm{cav}}$ and $6t_{\rm{syn}}=t_{\rm{cav}}$. Uncertainties in the magnetic field strengths and break frequencies may be the cause of this wide range. In a strong field, particles lose energy rapidly, but in a weak field the decay of the particles can be significantly slower. As a consequence, the particles can be much older than they look; in such cases the synchrotron age is a lower limit on the true age. Additionally, some of the systems may be in a driving state (the cavities are continuously pumped by the radio source) in which the buoyancy assumption doesn't apply.

On the other hand, A262\index{A262} and A478\index{A478} are completely different from the rest of the objects. In these cases the synchrotron age is higher than the X-ray age by a factor of 5. It is possible that the X-ray age was underestimated due to projection effects, but that would account for only a factor of $\sqrt{2}$ on average. A more plausible explanation for these cases is that the break frequency is in error. A262\index{A262} and A478\index{A478} are two out of three objects for which we have lobe emission at only 2 frequencies (327 MHz and 1.4 GHz). In these cases the break frequency was estimated using the method of \cite{myer85}. Therefore, break frequency and magnetic field misestimates may be the reasons for the discrepancy between the synchrotron ages and the buoyancy ages.

\subsection{Particle Content}
We calculated the ratio of the proton energy to electron energy ($k$), assuming $\phi=1$ for three different scenarios: that equipartition applies, that the synchrotron age equals the buoyancy age, and that the synchrotron age equals the sound speed age. In each scenario, we first used the volume from the X-ray maps (cavity volume) and then repeated the calculations using the volume from the 327 MHz radio maps (lobe volume). From theoretical estimates, $k$ is predicted to range between 1 and 2000, depending on the mechanism generating the electrons \cite{pach70}. The typical values used in the literature are $k=1$ for an electron-positron plasma and $k=100$ from cosmic rays assumptions. We found that for most of the cavities, $k$ is on the order of hundreds, but there  is a wide range from a few to a few thousand or more. Our range is similar to that of \cite{fabi02, dunn04, dunn05}, although our analyses differ in details. Such high values of  $k$ imply that in these systems other particles beside electrons are provinding additional pressure support, for example protons \cite{deya06}.

\begin{figure}
\includegraphics[scale=.5]{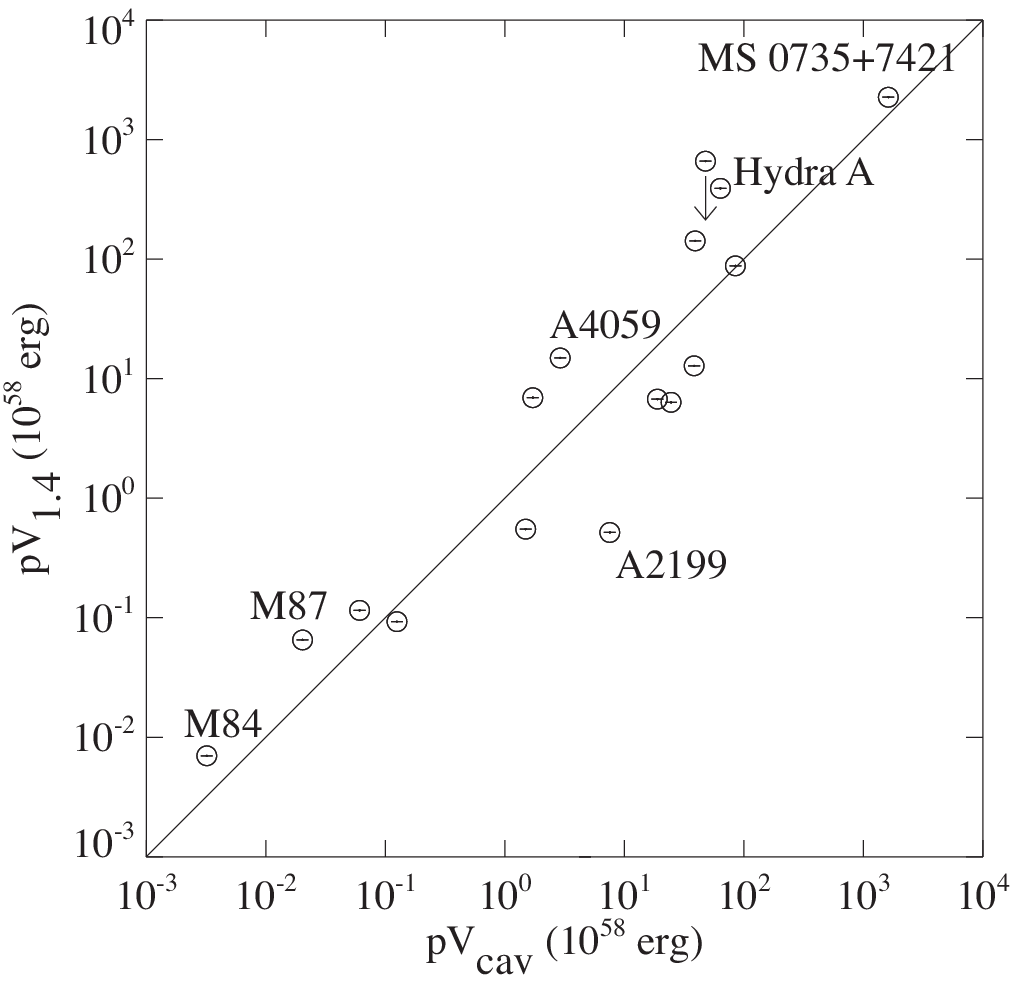}
\includegraphics[scale=.5]{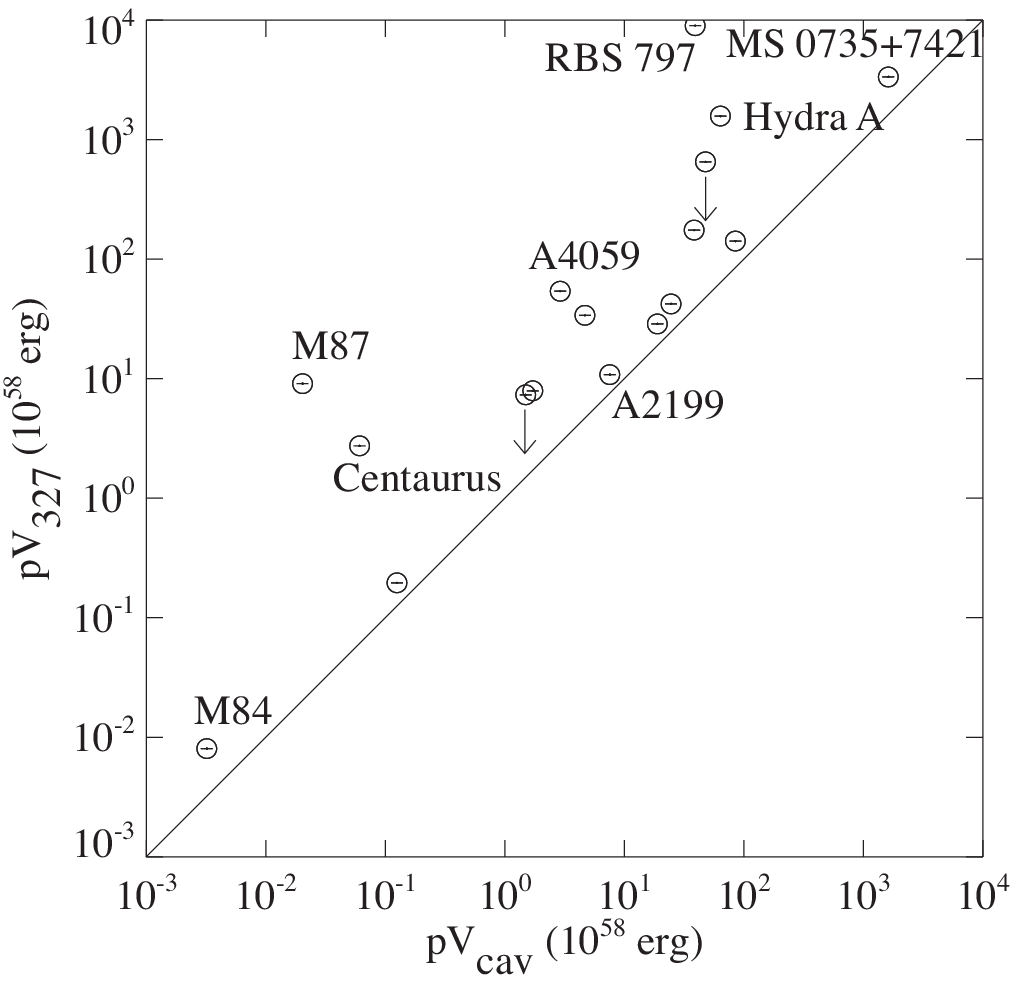}
\caption{\emph{Left}: $pV_{\rm{1.4}}$ versus $pV_{\rm{cav}}$. For Hydra A, $V_{\rm{1.4}}$ is calculated from the radio map of \cite{lane04}.  \emph{Right}: $pV_{\rm{327}}$ versus $pV_{\rm{cav}}$. For M87\index{M87} and Hydra A\index{HydraA}, $V_{\rm{327}}$ is calculated from the maps of \cite{owen00} and \cite{lane04}, respectively. The upper limits in both plots denote unresolved radio sources.}
\end{figure} 

\subsection{Radio Observations as a Tracer of Cavity Sizes }

Because synchrotron losses are smaller at 327 MHz than at higher frequencies, 327 MHz radio maps have proved to be crucial for detecting the outer, fainter cavities in Hydra A  \cite{wise06, lane04}. In order to evaluate whether 327 MHz emission is generally a better predictor of cavity size than 1.4 GHz emission, we compare $pV_{\rm{cav}}$, where $V_{\rm{cav}}$ is the volume of the cavities from the X-ray maps against $pV_{\rm{1.4}}$, where $V_{\rm{1.4}}$ is the volume of the lobes from 1.4 GHz radio maps, in Figure 2, left. We also plot $pV_{\rm{cav}}$ against $pV_{\rm{327}}$ (Figure 2, right). The pressure on each axes is measured at the position of the radio lobe's center or cavity's center, and can therefore differ from one axis to another. 

Figure 2 suggests that 1.4 GHz emission is a good tracer of cavity size in most cases, but there is a significant scatter, with many points falling below the equality line. Assuming that $P_{\rm{cav}}$ traces the minimum jet power, we note that 1.4 GHz underestimates the cavity power in some systems. On the other hand, at 327 MHz all of the points are above the line. It appears that 327 MHz recovers the energy better and is more sensitive to extended emission than 1.4 GHz. 327 MHz can be used as a predictor for the existence of X-ray cavities, such as in those objects that are well above the equality line in Figure 2, right (e.g., Centaurus\index{Centaurus} and RBS 797\index{RBS797}). In both plots, some of the objects are on the equality line (e.g., M84\index{M84} and MS 0735+74\index{MS0735+7421}). In these cases, both 327 MHz and 1.4 GHz emission fills the cavities, and both frequencies are good tracers of cavity size. However, some of the objects that at 1.4 GHz lie below the equality line lie on the line at 327 MHz (e.g., A2199\index{A2199}). In these cases, 327 MHz emission fills the cavities more fully. Therefore, for these objects 327 MHz emission is a better tracer of cavity size than 1.4 GHz emission.

Other objects are above the equality line in both plots (e.g., Abell 4059\index{A4059}, Hydra A\index{HydraA} and M87\index{M87}). In these cases, the radio lobes are larger than the cavities. It is possible that the X-ray measurements underestimated the size of these cavities or that the radio plasma has diffused beyond the cavities.

In Hydra A\index{HydraA}, the inner cavities that we measured from X-ray images are filled with 4.5 GHz radio emission. However, from  \cite{wise06} we know that Hydra A\index{HydraA} has outer cavities that are filled with 327 MHz radio emission \cite{lane04}. With this information in mind we note that in Hydra A\index{HydraA},  $pV_{\rm{cav}}$ will increase when we add the contribution from these outer cavities. A similar situation is true in M87\index{M87}, where  \cite{form06} discovered large outer cavities filled with 327 MHz radio emission  \cite{owen00}. There are other objects similar to Hydra A\index{HydraA} where the $pV$ inferred from the 327 MHz radio emission is well above the $pV$ we measure from X-ray images (e.g., Centaurus\index{Centaurus} and RBS 797\index{RBS797}). In these objects we expect to see outer cavities in deeper Chandra images which will increase their total $pV_{\rm{cav}}$ estimates.
  
\section{Conclusions}

We have presented an analysis of new, high-resolution radio data for a sample of 16 systems with X-ray cavities located in cluster cores. By combining X-ray data and VLA radio data taken at multiple frequencies, we find that the radiative efficiency in these systems is $\sim$1\%, but can be much lower. We find that the synchrotron ages of the lobes are generally less than the X-ray ages of the cavities, suggesting that the cavities are being pumped. We find that most of the lobes are not in equipartition. We conclude from the high ratios of proton to electron energy that other particles besides electrons are needed to provide additional pressure support. By comparing the size of the cavities from the X-ray data with the size of the lobes from the radio maps, we conclude that  327 MHz maps are a better tool than higher frequency maps for studying the history of AGN activity in the cores of clusters over the past several hundred million years. Finally, we note that 327 MHz radio maps can be a good proxy for X-ray cavities in systems where it is difficult to image the cavities directly in X-rays.

\end{document}